\documentclass[aps,prb,twocolumn,groupedaddress]{revtex4}

\usepackage[version=3]{mhchem} 
\usepackage{multirow}
\usepackage{graphicx}

\newcommand*{\density}[4]{$#1\times10^{#2} /$ #3$^{#4}$}

\begin{document}

\title{Triangulating tunneling resonances in a point contact}

\author{Nathaniel C. Bishop}
\email{nbishop@sandia.gov}
\affiliation{Sandia National Laboratories, Albuquerque, NM, USA}
\author{Carlos \surname{Borr\'as Pinilla}}
\altaffiliation{Universidad Industrial de Santander-Colombia}
\affiliation{University of Oklahoma, Norman, OK, USA}
\author{Harold L. Stalford}
\affiliation{University of Oklahoma, Norman, OK, USA}
\author{Ralph W. Young}
\affiliation{Sandia National Laboratories, Albuquerque, NM, USA}
\author{Gregory A. Ten Eyck}
\affiliation{Sandia National Laboratories, Albuquerque, NM, USA}
\author{Joel R. Wendt}
\affiliation{Sandia National Laboratories, Albuquerque, NM, USA}
\author{Edward S. Bielejec}
\affiliation{Sandia National Laboratories, Albuquerque, NM, USA}
\author{Kevin Eng}
\altaffiliation{Current affiliation: HRL Laboratories, Malibu, CA, USA}
\affiliation{Sandia National Laboratories, Albuquerque, NM, USA}
\author{Lisa A. Tracy}
\affiliation{Sandia National Laboratories, Albuquerque, NM, USA}
\author{Michael P. Lilly}
\affiliation{Sandia National Laboratories, Albuquerque, NM, USA}
\author{Malcolm S. Carroll}
\affiliation{Sandia National Laboratories, Albuquerque, NM, USA}

\date{\today}

\begin{abstract}
  We observe resonant tunneling in silicon split gate point contacts implanted with antimony and defined in a self-aligned poly-silicon double gate enhancement mode Si-MOS device structure.  We identify which resonances are likely candidates for transport through the antimony donor as opposed to unintentional disorder induced potentials using capacitance triangulation.  We determine the capacitances from the resonant feature to each of the conducting gates and the source/drain two dimensional electron gas regions.  In our device geometry, these capacitances provide information about the resonance location in three dimensions.  Semi-classical electrostatic simulations of capacitance, already used to map quantum dot size and position, identify a combination of location and confinement potential size that satisfy our experimental observations.  The sensitivity of simulation to position and size allow us to triangulate possible locations of the resonant level with nanometer resolution.  We discuss our results and how they may apply to resonant tunneling through a single donor.
\end{abstract}

\pacs{}

\maketitle

\section{Introduction}
Numerous groups are developing single electron spin devices in silicon\cite{SimmonsPRL2011, MorelloNature2010, XiaoPRL2010, LansbergenNatPhys2008}.  Electron spins in silicon can have very long coherence times\cite{McCameyScience2010, MortonNature2008}, which is desirable for both the basic study of spin interactions using engineered few spin systems as well as the pursuit of applications such as quantum computing and spintronics.  Donor atoms provide a natural confinement potential for single electrons at low temperature.  Donors placed intentionally (i.e., implant) or adventitiously into tunnel barriers of silicon devices have been probed using transport spectroscopy and charge sensing\cite{MorelloNature2010, LansbergenNatPhys2008, TanNanoLett2010}.  Single electron spin measurements have been reported recently in intentionally implanted structures introducing a tantalizing fabrication path towards future single donor spin devices\cite{MorelloNature2010, TanNanoLett2010}.  Two common challenges to developing single donor devices are establishing strict control over the position of the donor and avoiding interference from defects.

In this letter we report fabrication, measurement and modeling of a self-aligned, donor-implanted nanostructure that provides a path towards an improved alignment of donors to neighboring ohmic leads and lateral charge sensor structures.  Measurements of implanted devices show more tunneling resonances in implanted tunnel barriers, with respect to unimplanted tunnel barriers.  The resonant tunneling locations are triangulated using measured capacitance ratios.  Triangulation reveals that some resonances likely correspond to transport through regions that are not intentionally implanted.  This highlights the value of geometrical analysis, such as triangulation, to improve the confidence in identification of resonances that are candidates for transport through intentionally implanted donors instead of defects.  The technique of triangulation is, furthermore, more generally applicable to any system for which the presence of defects introduces ambiguity about the source/location of single electron behavior, for example, silicon, graphene and carbon nanotubes.

\section{Experiment}
\begin{figure}
\includegraphics[width=80mm]{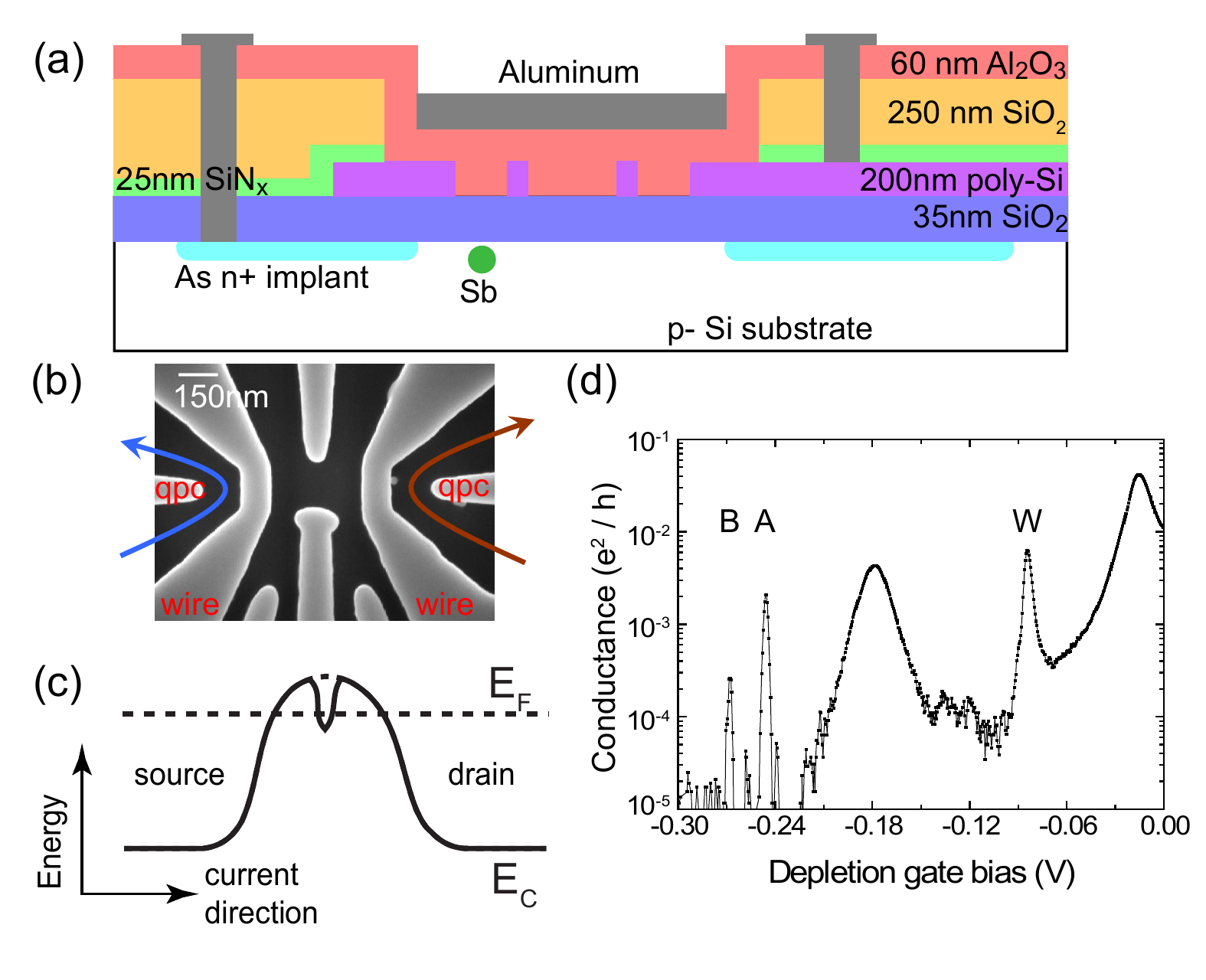}
\caption{(color online) (a) Cross section of our device, not to scale, showing polysilicon depletion gates and aluminum inversion gate.  (b) Scanning electron micrograph of the polysilicon depletion gate layout, from a representative sample.  (c) Schematic energy diagram, along the blue arrow in part (b).  (d) Conductance through the left constriction (blue arrow in part (b)) at $T = 20$ mK, showing sub-threshold tunneling resonances (labeled A, B, and W).\label{figure1}}
\end{figure}

We fabricated and measured a point contact formed by a split gate geometry in an enhancement mode Si-MOS nanostructure, Fig. \ref{figure1}(a).  Poly-silicon (poly), deposited 200 nm thick and then degenerately doped via ion implantation, form first level gates, used to deplete the electrons and form the point contact.  They are isolated from the silicon substrate by a 35 nm thermal \ce{SiO2} layer.  The second level, an enhancement gate, is formed with a 100 nm aluminum layer, covering the entire active region of the device and overlapping highly doped substrate regions which serve as ohmic contacts to the induced electron layer.  The Al top gate is isolated from the poly by a 60 nm aluminum oxide layer grown by atomic layer deposition (ALD).  Quantum dots with periodic Coulomb blockade and lateral charge sensing of the quantum dot electron occupation have been demonstrated with similar processing\cite{NordbergPRB2009, NordbergAPL2009}.

A small number of antimony atoms were implanted in to one of two point contacts in the device.  After the poly gate pattern is etched, Fig. \ref{figure1}(b), the device was covered with 300 nm of PMMA, and an 80x80 nm$^{2}$ opening was formed in one constriction (left side, blue arrow), while a second constriction remained unexposed (right side, brown arrow).   A blanket 100 keV Sb implant with dose of \density{4}{11}{cm}{2} was used.  SRIM simulations\cite{SRIM} project that the combined thickness of the 35 nm gate oxide and the PMMA should result in Sb concentrations below the p-type background of the substrate, $>100 \ $Ohm-cm or $\sim10^{14} / $cm$^{3}$.  The dose and energy correspond to an average of 8 Sb ions in the 80x80 nm$^{2}$ window, at a depth of 10 nm below the Si/\ce{SiO2} interface, according to simulations done with SRIM.  Roughly 50\% of the dose is deposited in the \ce{SiO2} dielectric.  The implanted constriction acts as a control case, to characterize non-implant process damage that might mimic donor assisted tunneling through  the split gate barrier.

The device structure is completed by reoxidation at $900\ ^{0}$C for 24 minutes followed by the addition of ALD \ce{Al2O3} and Al metallization, as described previously\cite{NordbergAPL2009}.  The reoxidation step activates the dopant and anneals damage.  High frequency and quasi-static capacitance-voltage measurements were done to characterize the residual damage for a similar process flow using blanket depositions, implants and large Al capacitor structures, shown in Table \ref{table1}.  To within the uncertainty of the measurement, reoxidation removes residual damage from implant whereas a rapid thermal anneal leaves non-negligible defect charge in the oxide, which could introduce a higher number of defect related resonances.  A simple estimate for the diffusion length of Sb for this thermal budget is 6 nm\cite{FairJMR1986}.  We note that because Sb is a vacancy based diffuser, excess interstitials introduced due to implant and oxidation that enhance other dopant diffusion will not greatly enhance Sb diffusion, and may suppress it\cite{AlzankiSST2004}.

In Fig. \ref{figure1}(c), we plot a schematic model of the electron potential in the Si channel, along the blue curve in part (b).  At the extremes are high-density source and drain regions, where the Fermi energy is well above the conduction band edge.  Between the two is a depletion region, where the negative bias on the depletion gates has brought the conduction band edge above the Fermi energy, and brought an energy level of a hypothetical donor or trap below the band edge into resonance with the source and drain.

We examine the point contacts using transport spectroscopy.  The threshold voltage through the implanted point contact is appreciably higher than the control, 17 and 10 volts, respectively.  To determine the threshold, we extrapolate to zero conductance from a region where the conductance is linearly related to the enhancement gate bias.  \ref{figure1} (d) shows the current through the implanted point contact (left side, blue arrow in Fig. \ref{figure1}(b)) as a function of depletion gate bias.  We measure the conductance using low-frequency lock-in techniques, in this case a $15 - 100 \mu$ Vrms AC excitation at 27 Hz, and measured the resulting current.  

We observed numerous peaks in the sub-threshold conductance, in contrast to a single small peak observed in the control side.  Several other samples show a similar trend of increased numbers of resonances on the Sb implanted point contact compared to controls.   A point implanted with silicon, dose \density{4}{11}{cm}{2} and energy 45 keV, chosen for its similar distribution of Si interstitials as the Sb implants, did not show a substantial increase in resonances below threshold.  Sub-threshold conductance resonances through a point contact normally result from tunneling through a localized confinement potential within the tunnel barrier.  The confinement can be due to intentionally implanted donor potentials or unintentional damage related defects (e.g., charged defects in the oxide) or impurities (e.g., metals or other background dopants).  We will discuss results primarily from the resonances labeled A, B, and W.  

\begin{table}
\caption{Device fabrication process damage parameters, from capacitance vs. voltage measurements on planar test devices.\label{table1}}
\begin{tabular}{l | c | c}
\multirow{2}{*}{Sample type} & $Q_{fb}$ & D$_{it}$  \\ 
					& (q/cm$^{2}$)  & (q/cm$^{2}$eV) \\ \hline
Normal process & $+5\times10^{10}$ & $1\times10^{10}$ \\ \hline
Silicon implant + 30 sec @ 800 C & \multirow{2}{*}{$+3\times10^{11}$} & \multirow{2}{*}{$2\times10^{11}$} \\
(Donor device equivalent dose)      & & \\ \hline
Silicon implant + 24 min @ 900 C & $+1\times10^{11}$ & $1\times10^{10}$
\end{tabular}
\end{table}

\begin{figure}
\includegraphics[width=80mm]{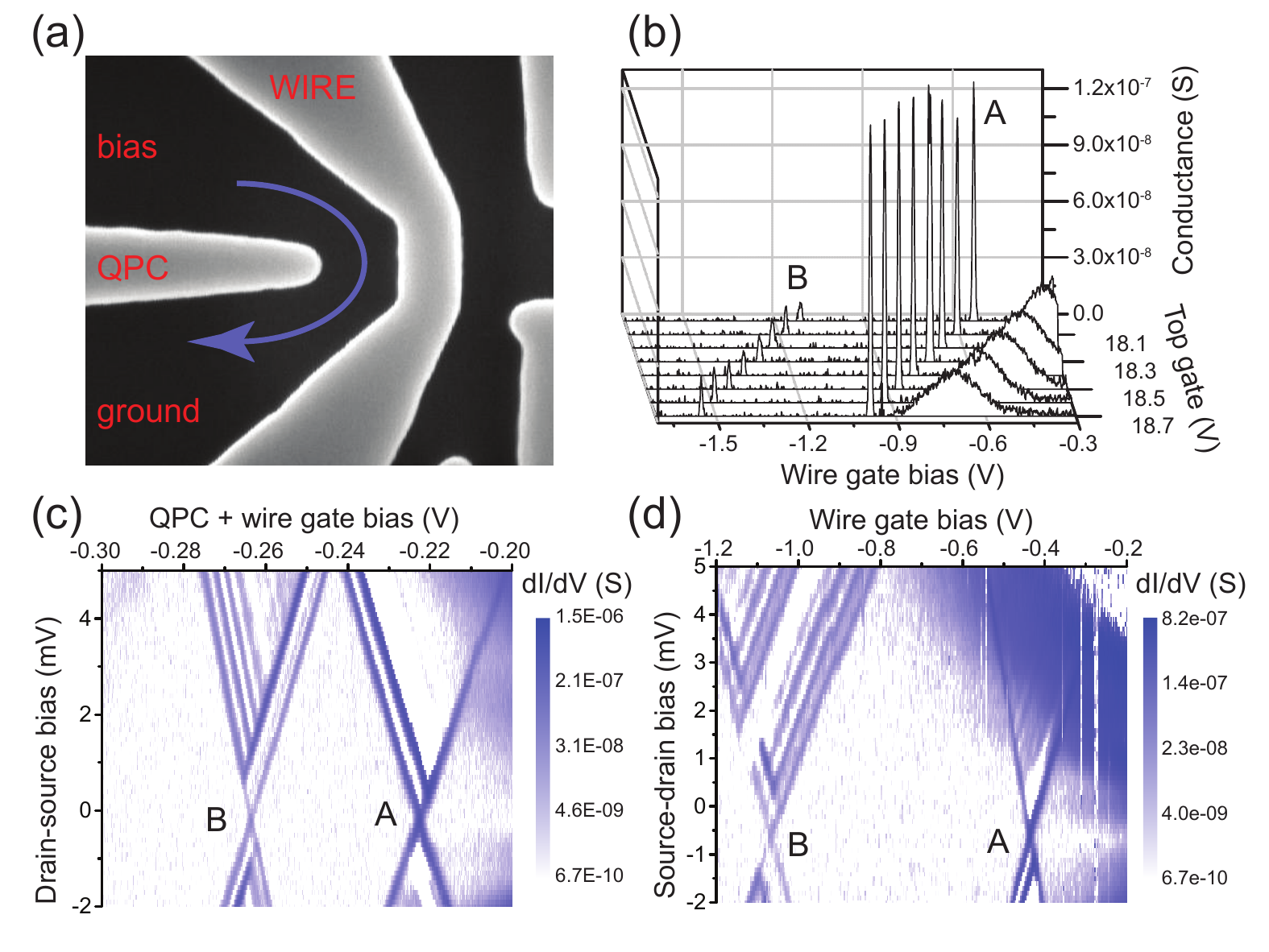}
\caption{(a) Close up SEM image of the implanted constriction, with labels for the depletion gates used in this analysis.  (b) Conductance as a function of depletion gate bias and top (aluminum) gate bias.  The movement of peaks A and B, $dV_{wire}/dV_{TG}$, is determined by the ratio of capacitances $C_{top} / C_{wire}$.  (c)  Conductance vs. source-drain bias and gate bias (both qpc and wire gates).  Source grounded.  (d) Conductance vs. source-drain bias and gate bias (wire gate only, qpc fixed).  Drain grounded.\label{figure2}}
\end{figure}

We measured the tunneling current dependence on each gate bias, as well as the source-drain bias across the constriction.  Several sets of data, each taken with a different configuration, are shown in Fig. \ref{figure2}(b - d).  Peaks A, B, and W (not shown here) all exhibit clearly defined triangular regions of high conductance consistent with discrete electron, Coulomb blockaded transport.  If the drain 2DEG region is held at ground, the slope of the positive-going line demarcating the high-conductance region is a measure of the ratio of capacitances $C_{gate} / (C - C_{S})$.  Here $C$ is the total capacitance from the localized electron state to all conductors in the device, $C_{S}$ is the capacitance between the source 2DEG region and the localized state, and $C_{gate}$ is the capacitance between the depletion gate and the localized state.  The slope of the negative-going line is a measure of the ratio $C_{gate} / C_{S}$.  By performing this measurement for both the wire and qpc gates, and by switching which 2DEG region is held at ground potential, we obtained a set of capacitance ratios for each peak.  A set of measurements for the constriction determines $C_{S} / C$, $C_{D} / C$, $C_{wire} / C$, $C_{qpc} / C$, and $C_{top} / C$, which together account for more than $97\%$ of the total capacitance (Fig. \ref{figure2}(c) and (d)).  To determine the top (aluminum) gate capacitance ratio, we measure the position of each peak as a function of top gate and (e.g.) wire gate voltage, which determines the ratio $C_{top} / C_{wire}$.   Note that for the measurements of peak W, an incomplete set of data allows us to only determine the total $C_{wire} + C_{qpc} / C$, but we cannot determine separate values for $C_{wire} / C$ or $C_{qpc} / C$.

We note that the measured capacitance ratios are found to be self-consistent.  That is, a capacitance ratio such as $C_{D} / C$ can be directly measured or indirectly estimated through measurement of the ratios $C_{S} / C$, $C_{qpc} / C$, $C_{wire} / C$, and $C_{top} / C$, in conjunction with the constraint that the sum of all capacitance ratios must equal unity.  All methods of determining each ratio are in good agreement with one another to within $5\%$, despite uncertainty due to drift from measurement to measurement, and difficulty in precision determination of the slope within a single measurement.

Assuming the capacitance is primarily a geometric effect, these ratios contain information about the spatial location of the bound electrons.  Different ratios suggest different spatial position or size of the confined electron at each resonance.  Measuring $C_{wire} / C$ and $C_{qpc} / C$ determines the position of the localized state perpendicular to the current direction, while $C_{S} / C$ and $C_{D} / C$ determine the position along the current direction.  We have tabulated the results for each of the three points of interest in Table \ref{table2}.

\begin{table*}
\caption{Measured vs. simulated capacitance ratios for three tunnel resonances in our device.\label{table2}}
\begin{tabular}{c | c | c | c | c | c | c}
\multirow{2}{*}{} & \multicolumn{2}{c |}{B} & \multicolumn{2}{| c |}{A} & \multicolumn{2}{| c}{W} \\
 & Expt. & Sim. & Expt. & Sim. & Expt. & Sim. \\ \hline
$C_{S} / C$ & $0.436 \pm 0.018$ & $0.435$ & $0.378 \pm 0.014$ & $0.379$ & $0.257 \pm 0.024$ & $0.257$ \\ \hline
$C_{D} / C$ & $0.393 \pm 0.010$ & $0.392$ & $0.474 \pm 0.006$ & $0.473$ & $0.627 \pm 0.053$ & $0.626$ \\ \hline
$C_{wire} / C$ & $0.014 \pm 0.013$ & $0.013$ & $0.019 \pm 0.001$ & $0.019$ & \multirow{2}{*}{$0.094 \pm 0.009$} & $0.0025$ \\ \cline{1-5} \cline{7-7}
$C_{QPC} / C$ & $0.134 \pm 0.014$ & $0.132$ & $0.099 \pm 0.007$ & $0.095$ &  & $0.060$ \\
\end{tabular}
\end{table*}

\section{Modeling}
\begin{figure}
\includegraphics[width=80mm]{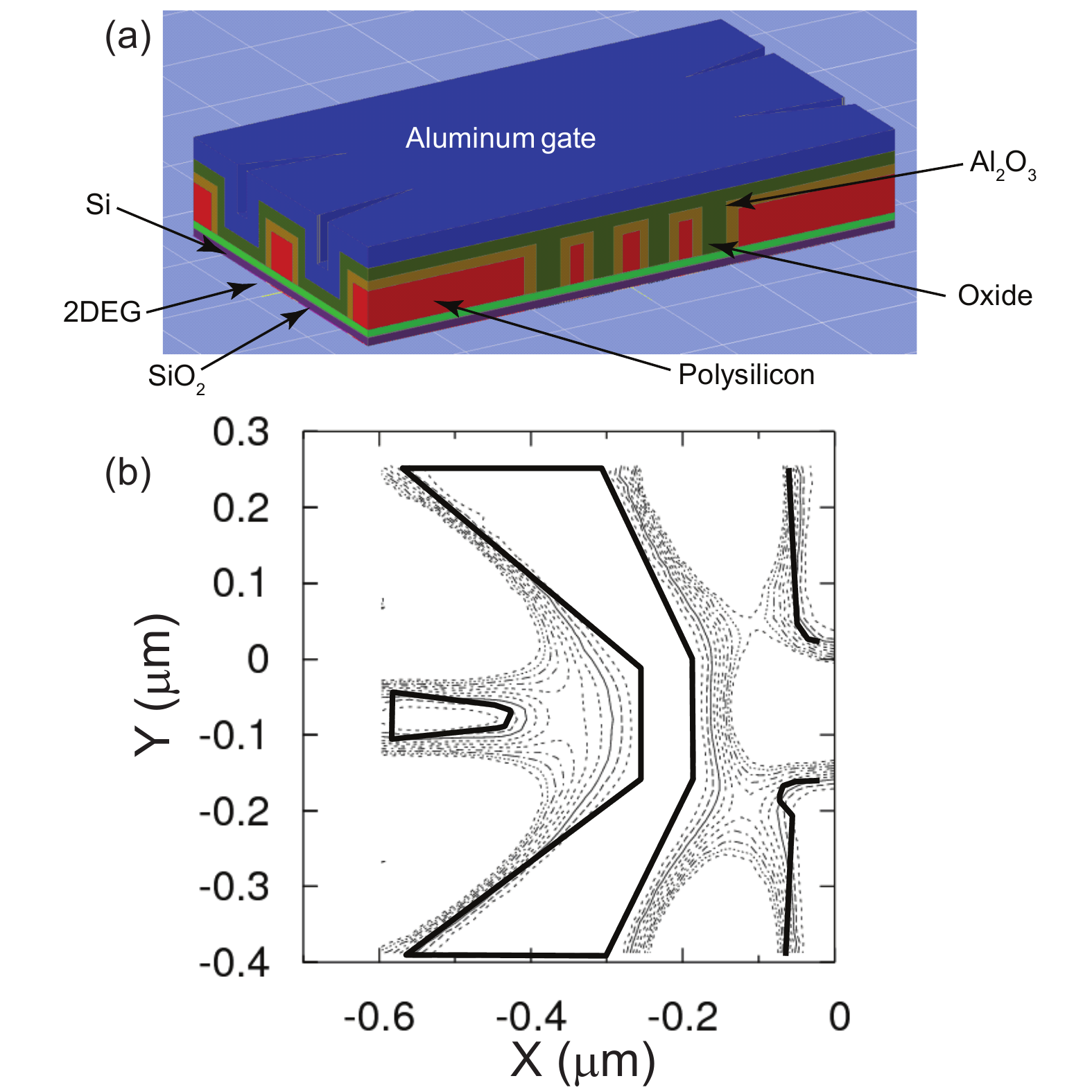}
\caption{(a) An oblique view of the 3D structure of our device, used for calculating the 2D electron density in the channel, shown in (b).  (b) Constant electron density contours, from \density{1}{11}{cm}{2} to \density{1}{12}{cm}{2}, in steps of \density{1}{11}{cm}{2}.  We take the size of the source and drain regions to be that area with density higher than \density{1.5}{11}{cm}{2}.\label{figure3}}
\end{figure}

\begin{figure}
\includegraphics[width=80mm]{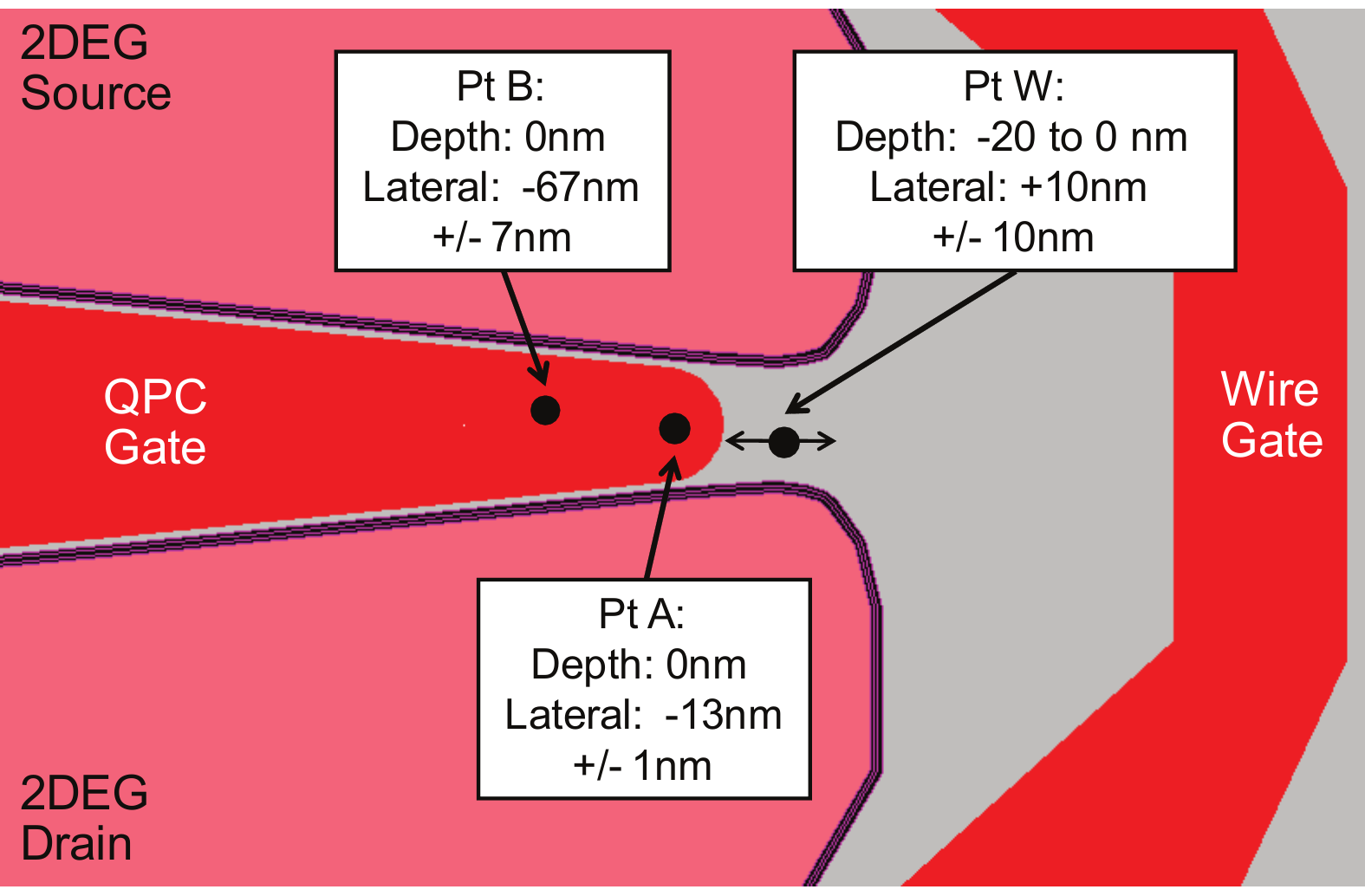}
\caption{Best fit locations for resonances A, B, and W.  The simulation determines a unique position for each possible combination of measurements.  These error bars reflect experimental error, and do not include uncertainty from possible disagreement between the model and the actual device, which is expected to be small.\label{figure4}}
\end{figure}

The measured capacitance ratios bound the possible location of a single electron on a donor through the geometric dependences of the capacitance to position.   A detailed capacitance model is used to establish quantitatively the predicted position (i.e., capacitance triangulation) and confidence in the triangulation.   A semi-classical Thomas-Fermi calculation done in TCAD Sentaurus\cite{sentaurus} is used to establish the position and density of the 2DEG.  Then capacitances are calculated using the commercial finite element modeling package CFD-ACE+\cite{cfdace}, in which all poly gates, the top Al gate and the S/D 2DEGs are modeled as perfect conductors.   A 2DEG density of \density{1.5}{11}{cm}{2} is used as the metallic edge for the 2DEG.  Any density above \density{1.5}{11}{cm}{2} we regard as perfectly conducting, and below as perfectly insulating.  This value is similar to an experimentally measured metal-insulator-transition temperature for MOS 2DEGs\cite{TracyPRB2009} and has been successfully used in this way for modeling quantum dot capacitances\cite{StalfordIEEENano2010}.  The capacitances between the gates and a 1.8 nm radius sphere, the Bohr radius of the first bound electron on antimony in bulk silicon, are calculated in this case to examine what antimony locations would be consistent with the measurements.  \ref{figure3}(a) is a view of the 3D model used for this calculation, and a contour plot of electron density for the gate biases used in the measurements of \ref{figure2}(b) is shown in \ref{figure3}(b).  A diagram of all the conductors in the simulation, labeled C1 to C19, is shown in \ref{figure3}(c).  Conductors C3 and C4 represent a single trial position for a donor in the tunnel barrier regions.  We chose positions by trial and error until a good match to the measured capacitance ratios was found.  Further refinement of the position was established through interpolation between the database of previous positions and capacitances, \ref{figure4}.  The interpolation produces an optimal single position for each resonance.  Negative values indicate a position under the poly gate (i.e., an unlikely location for an intentionally implanted Sb).  The W peak position, triangulated by capacitance ratios, is the only resonance that is consistent with transport through a location between the poly split gates where Sb was targeted.  
 
The capacitance model indicates that small differences in capacitance ratios can correspond to significant positional shifts of many nanometers.  For example, the measured capacitances ratios ($C_{wire} / C$ and $C_{QPC} / C$) for Pts. A and B differ by 24-36\%; the resulting lateral positional shift (along the QPC gate) from Pt. B to Pt. A is $54\ $ nm as determined by simulation, \ref{figure4}.  For the source and drain regions, Pt. B has $C_{S}$ 5\% larger than $C_{D}$, Pt. A has $C_{D}$ 11\% larger than $C_{S}$; the resulting perpendicular positional shift (from source to drain) for Pt. B is $+2\ $ nm and $-4\ $ nm for Pt. A.  This highlights the potential for high precision triangulation using this method, but also highlights the potential for error due to uncertainties in the exact gate dimensions.  Lateral dimensions of the poly are measured by scanning electron microscopy on a representative sample.  We estimate that the uncertainty in dimensions is dominated by sample to sample variation, and is less than or equal to $15\ $ nm for gaps and widths.
 
\section{Discussion}
In three dimensions, we can determine lateral position relatively easily (position parallel to the silicon-oxide interface), but vertical displacement is less direct.  The capacitance from source or drain to the localized state determines both the length of the tunneling region that forms in the constriction, as well as any displacement from the center of the tunnel barrier.  If we assume that the source and drain regions have similar shapes, any difference in source vs. drain capacitance must be due to displacement from the mid-point between the source and drain in the tunnel barrier.  As tunneling rates are exponentially sensitive to tunnel barrier length, we would not expect to measure significant displacements from the mid-point, since the longer tunnel barrier will limit transport, and quickly reduce the current below our noise floor.  This is consistent with the triangulated positions that are estimated to have displacements of $5$ nm or less from the center, for tunnel barriers of $40 - 50$ nm total length.

The lateral position is constrained by the relative capacitive coupling of the wire vs. the QPC gate.  We find a much stronger than expected coupling from the QPC gate, especially for peaks A and B, where it dominates the coupling from the wire gate by a factor of 5 or 10 times.  This ratio pushes the triangulated position far from the wire gate, despite its large area.  For peaks A and B, the simulated positions which best fit the data are found below the QPC gate.  Even the incomplete set of data for peak W allows us to determine its lateral position, though with a larger error bar.  Due to the changing length of the tunnel barriers as a fuction of lateral position, \textit{i.e.} the distance between source and drain, easily seen in \ref{figure4}, the magnitude of $C_{S} / C$ and $C_{D} / C$ also depend on lateral position, providing a position for peak W.

The lateral position of all the resonances are also forced closer to or further under the QPC gate in order to produce a sufficiently small $C_{S} / C$ ratio.  This is driven by the gap between the 2DEGs, which expands out as the QPC gate widens.  The lateral position below the QPC is, therefore, influenced significantly by relatively small changes in the edge of the 2DEG, which is one of the primary sources of lateral position uncertainty under the QPC since the 2DEG position is only estimated through the semi-classical calculation and the assumed MIT metallic boundary cut-off.  However, even if the 2DEG is more depleted than indicated by the semi-classical calculation, leading to wider gaps, the donor position would still necessarily be under the QPC gate because of the $C_{wire} / C_{QPC}$ ratio.

The magnitude of confidence in position provided by this triangulation method is sensitive to several factors including uncertainty in exact dimensions (e.g., imperfect knowledge about device dimensions) discussed earlier as well as three other underlying assumptions discussed now.  First, the calculation assumes that the Sb electron wavefunction radius is 1.8 nm, which implies that the electron wavefunction is bulk like.  Donors near the surface in the presence of vertical electric fields are predicted to hybridize to a bound surface state\cite{LansbergenNatPhys2008}, which can considerably increase the size of the electron distribution and shift it to the surface.  Second, the calculations assume the radius of a single electron occupation Sb site, although it is possible that the resonance corresponds to a $N = 1$ to $N = 2$ (D-) transition.  The size of the D- wavefunction is considerably larger, probably about $3\ $ nm radius.  Both of these cases represent circumstances where the calculations have underestimated the size of electron wavefunction.  Simulations indicate that an increase of radial size of order of 2 - 3 does not significantly change the position.  This highlights an insensitivity of using capacitance ratio triangulation to absolute size if it is relatively small compared to the tunnel barrier gap size.  If the total capacitance can be obtained through a charging energy measurement, then this also allows size determination, which was not available in these cases.  

Triangulation using the measured capacitance ratio leads to the conclusion that the only way that resonances A or B can be assigned to transport through the D+ to D0 transition of antimony sites is that the Sb is located below the poly gates.   It is unlikely that Sb is at these locations because the poly is 200 nm further masking this region, in addition to the PMMA, from direct implant and reasonable diffusion lengths for the thermal budget are small enough to make it very unlikely that Sb is in this region.  Defects near the Si/\ce{SiO2} interface are a likely source of a confining potential.  Gate oxides are known to exhibit point like positive charge defects near the silicon/oxide interface, which would produce relatively deep electrostatic confining wells\cite{NordbergAPL2009}.  Best gate oxides are reported to have fixed charge densities on the order of $10^{11} / $cm$^{2}$, making it probable that over the length and width of the QPC gate there will be 1 or 2 defects.  Faster interface state densities, Dit, can be smaller densities and are therefore less of a concern.  Although the confinement of the electron will not be the same as that of antimony, 1.8 nm Bohr radius, calculations of slightly larger metallic discs, to simulate surface disorder dots, are not inconsistent with the capacitance ratios.  A less likely possible source of the resonance could be arsenic dopants that diffused through the poly and gate oxide.  Simulations using a commercially available process simulator (TSUPREM4) indicate that the As concentration is well below the boron background concentration of $\sim10^{14} / $cm$^{3}$.  If true, it is unlikely that these resonances come from arsenic.  Implant and diffusion of arsenic is relatively well understood through years of calibrated modeling developed for the CMOS industry making simulated projections of dopant depth fairly reliable.

\section{Summary}
A tunnel barrier in a silicon split-gate point contact was implanted with a small number of Sb donors and characterized electrically at $T \simeq 300\ $ mK.  The device fabrication is self-aligned using poly depletion gates.  The self-aligned process flow represents a first time proof of principle path towards significantly better lateral control of donor placement within tunnel barriers and is compatible with single ion detectors\cite{BielejecNano2010}.  The implanted tunnel barrier shows considerably more single electron transport resonances than an unimplanted, control, case.  The ratio of the sum of capacitances to the gate capacitance was characterized for all available gates for three sharp resonances.  Triangulation of the positions of the resonance is made possible through the geometric dependence of the capacitances and the positioning of the conducting gates, which are configured so that all three spatial directions are probed.  A semi-classical capacitance model was used to bound likely regions that a 1.8 nm radius Sb electron would be consistent with the measured ratios.  One of the resonances is consistent with transport through a donor within the implanted constriction.  The triangulation method clearly indicates, on the other hand, that the other two resonances, if consistent with a small confined electron of order of 1.8 nm radius, would have located under the QPC gate.  Oxide related defects are a likely candidate that might facilitate transport under the poly gate although transport through adventitious background donors or other impurities cannot be ruled out.  The triangulation method therefore provides critical guidance to distinguish between likely defects and high probability candidates related to single electron transport through intentionally implanted donors.  It is furthermore a demonstration of a method to generally distinguish between intentionally engineered single electron potentials in contrast to unintentional resonances in more disordered systems.

\begin{acknowledgments}
This work was performed, in part, at the Center for Integrated Nanotechnologies, a U.S. DOE, Office of Basic Energy Sciences user facility. The work was supported by the Sandia National Laboratories Directed Research and Development Program. Sandia National Laboratories is a multi-program laboratory operated by Sandia Corporation, a Lockheed-Martin Company, for the U. S. Department of Energy under Contract No. DE-AC04-94AL85000.
\end{acknowledgments}

\bibliography{tunneltriangulation}

\end{document}